\documentclass[12pt]{article}
\usepackage{epsf, cite, amssymb}
\usepackage{epsfig}
\usepackage{amsmath}
\usepackage{latexsym}
\usepackage{epsfig}
\usepackage{cite}
\usepackage[english]{babel}
\usepackage{graphicx,color}
\usepackage[colorlinks=true, linkcolor=blue, bookmarks=true]{hyperref}

\newcommand{\arXiv}[1]{\href{http://www.arXiv.org/abs/#1}{#1}}

\makeatletter
\renewcommand\section{\@startsection {section}{1}{\z@}%
                               {-3.5ex \@plus -1ex \@minus -.2ex}
                               {2.3ex \@plus.2ex}%
                               {\normalfont\large\bfseries}}
\renewcommand\subsection{\@startsection{subsection}{2}{\z@}%
                                 {-3.25ex\@plus -1ex \@minus -.2ex}%
                                 {1.5ex \@plus .2ex}%
                                 {\normalfont\bfseries}}
\makeatother

\def\IZ{\relax\ifmmode\mathchoice
{\hbox{\cmss Z\kern-.4em Z}}{\hbox{\cmss Z\kern-.4em Z}}
{\lower.9pt\hbox{\cmsss Z\kern-.4em Z}} {\lower1.2pt\hbox{\cmsss
Z\kern-.4em Z}}\else{\cmss Z\kern-.4em Z}\fi}
\def\IR{\relax{\rm I\kern-.18em R}}

\def\one{{\hbox{ 1\kern-.8mm l}}}

\def\tr{{\rm tr\,}}

\newlength{\bredde}
\def\slash#1{\settowidth{\bredde}{$#1$}\ifmmode\,\raisebox{.15ex}{/}
\hspace*{-\bredde} #1\else$\,\raisebox{.15ex}{/}\hspace*{-\bredde}
#1$\fi}

\newsavebox{\zzzbar}
\sbox{\zzzbar}
{\setlength{\unitlength}{0.9em}
\begin{picture}(0.6,0.7)
\thinlines
\put(0,0){\line(1,0){0.6}}
\put(0,0.75){\line(1,0){0.575}}
\multiput(0,0)(0.0125,0.025){30}{\rule{0.3pt}{0.3pt}}
\multiput(0.2,0)(0.0125,0.025){30}{\rule{0.3pt}{0.3pt}}
\put(0,0.75){\line(0,-1){0.15}}
\put(0.015,0.75){\line(0,-1){0.1}}
\put(0.03,0.75){\line(0,-1){0.075}}
\put(0.045,0.75){\line(0,-1){0.05}}
\put(0.05,0.75){\line(0,-1){0.025}}
\put(0.6,0){\line(0,1){0.15}}
\put(0.585,0){\line(0,1){0.1}}
\put(0.57,0){\line(0,1){0.075}}
\put(0.555,0){\line(0,1){0.05}}
\put(0.55,0){\line(0,1){0.025}}
\end{picture}}

\newcommand{\ena}{\end{eqnarray}}
\newcommand{\beqa}{\begin{eqnarray}}
\newcommand{\eeqa}{\end{eqnarray}}
\newcommand{\bea}{\begin{eqnarray}}
\newcommand{\eea}{\end{eqnarray}}

\newcommand{\eq}[1]{(\ref{#1})}

\newcommand{\be}{\begin{equation}}
\newcommand{\ee}{\end{equation}}

\usepackage{graphicx}

\def\be{\begin{equation}}
\def\ee{\end{equation}}
\def\beq{\begin{eqnarray}}
\def\eeq{\end{eqnarray}}

\def\({\left (}
\def\){\right )}
\def\[{\left [}
\def\[{\right ]}

\def\tr{\mathrm{tr}}

\def\ba{\begin{eqnarray}}
\def\ea{\end{eqnarray}}

\input amssym.def
\input amssym.tex

\newcommand{\bbibitem}[1]{\bibitem{#1}\marginpar{#1}}
\def\Bibitem#1{\bibitem{#1}%
  \smash{\hbox to0pt{\raise1ex\hbox{\tiny[#1]}\hss}}}

\def\Label#1{\label{#1}%
  \smash{\hbox to0pt{\raise1ex\hbox{\tiny[#1]}\hss}}}
\def\noLabels{\let\Label=\label}
\def\nobbibitem{\let\bbibitem=\bibitem}
 \def\noBibitem{\let\Bibitem=\bibitem}

  \newcommand{\cN}{{\cal N}}

\newcommand{\beann}{\begin{eqnarray*}}  \newcommand{\eeann}{\end{eqnarray*}}
\newcommand{\bfig}{\begin{figure}} \newcommand{\efig}{\end{figure}}
\newcommand{\bcen}{\begin{center}} \newcommand{\ecen}{\end{center}}
\newcommand{\btab}{\begin{tabular}} \newcommand{\etab}{\end{tabular}}

\def\tr{\operatorname{tr\:}}

\newcommand{\vev}[1]{\left\langle{#1}\right\rangle}

\newtheorem{Proposition}{Proposition}[section]

\newtheorem{Theorem}{Theorem}[section]
\newtheorem{Lemma}{Lemma}[section]
\newtheorem{Corrolary}{Corrolary}[section]

\newcommand{\bp}{\begin{Proposition}}   \newcommand{\ep}{\end{Proposition}}
\newcommand{\bt}{\begin{Theorem}}   \newcommand{\et}{\end{Theorem}}
\newcommand{\bl}{\begin{Lemma}}     \newcommand{\el}{\end{Lemma}}
\newcommand{\bc}{\begin{Corrolary}} \newcommand{\ec}{\end{Corrolary}}
\begin{document}

\begin{titlepage}
\begin{flushright}
TAUP-2854/12\\
NSF-KITP-12-169
\end{flushright}
\vfill
\begin{center}
{\Large \bf Chern-Simons diffusion rate in a holographic Yang-Mills theory}

\vskip 10mm

{\large Ben Craps$^{a}$, Carlos Hoyos$^{b}$, Piotr Sur\'owka$^{a}$, Pieter Taels$^{a}$}

\vskip 7mm

$^a$Theoretische Natuurkunde, Vrije Universiteit Brussel, and\\ 
\hspace*{0.15cm}International Solvay Institutes, \\
\hspace*{0.15cm} Pleinlaan 2, B-1050 Brussels, Belgium. \\
$^b$Raymond and Beverly Sackler Faculty of Exact Sciences\\
School of Physics and Astronomy \\
\hspace*{0.15cm} Tel-Aviv University, Ramat-Aviv 69978, Israel.\\

\vskip 3mm
\vskip 3mm
{\small\noindent  {\tt Ben.Craps@vub.ac.be, choyos@post.tau.ac.il, Piotr.Surowka@vub.ac.be, Pieter.Taels@vub.ac.be}}

\end{center}
\vfill

\begin{center}
{\bf ABSTRACT}
\vspace{3mm}
\end{center}
Using holography, we compute the Chern-Simons diffusion rate of 4d gauge theories constructed by wrapping D4-branes on a circle. In the model with antiperiodic boundary conditions for fermions, we find that it scales like $T^6$ in the high-temperature phase. With periodic fermions, this scaling persists at low temperatures. The scaling is reminiscent of 6d hydrodynamic behavior even at temperatures small compared to compactification scales of the M5-branes from which the D4-branes descend. We offer a holographic explanation of this behavior by adding a new entry to the known map between D4 and M5 hydrodynamics, and suggest a field theory explanation based on ``deconstruction'' or ``fractionization''.
\vfill


\end{titlepage}


\section{Introduction}\label{intro}

The classical vacua of a non-Abelian gauge theory are labeled by an integer topological charge, the Chern-Simons number. Quantum tunneling (instantons) as well as thermal processes (sphalerons) can change the Chern-Simons number:
\be\label{NCS}
\Delta N_{CS}=\frac{1}{64\pi^2}\int d^4x \epsilon^{\mu\nu\rho\sigma} \, F^a_{\mu\nu} F^a_{\rho\sigma}.
\ee
The corresponding Chern-Simons diffusion rate is defined as
\be\label{GammaCS}
\Gamma_{CS}=\frac{\langle (\Delta N_{CS})^2\rangle}{Vt}= \int d^4x\left\langle  \frac{1}{64\pi^2} \epsilon^{\mu\nu\rho\sigma}\, F^a_{\mu\nu} F^a_{\rho\sigma}(x)\;   \frac{1}{64\pi^2} \epsilon^{\alpha\beta\gamma\delta}\, F^b_{\alpha\beta} F^b_{\gamma\delta}(0) \right\rangle,
\ee
where $Vt$ is the volume of space-time. This quantity is of great interest in the context of the chiral magnetic effect in heavy ion collisions \cite{Kharzeev:2004ey,Kharzeev:2007tn,Kharzeev:2007jp,Fukushima:2008xe}. Event-by-event fluctuations of the topological charge can produce an imbalance between left- and right- handed quarks in the quark-gluon plasma phase, generating a net axial charge. Then, in non-central collisions, strong magnetic fields can produce an electric current with a magnitude proportional to the axial charge, an effect due to the chiral anomaly. From the experimental point of view this could lead to a measurable separation of charge in the final products of the collisions. The Chern-Simons diffusion coefficient is also important for models of electroweak baryogenesis (see e.g. \cite{Cohen:1993nk}), since it is related to baryon number violation in the standard model. At weak coupling it takes the form \cite{Bodeker:1998hm,Bodeker:1999gx}, 
\be
\Gamma_{CS}^{YM}\propto g_{YM}^{10}\log(1/g_{YM}^2)\,T^4, \ \ \ \ \ {\rm for}\ SU(2),\  g_{YM}^2\ll 1.
\ee
At strong coupling it has been computed for $\cN=4$ super Yang-Mills (SYM) at  large $N$ and strong coupling, using holographic methods \cite{Son:2002sd}:
\be\label{GCSSYM}
\Gamma_{CS}^{SYM}=\frac{\lambda^2}{256 \pi^3}\,T^4, \ \ \ \ \ {\rm for}\ \lambda\equiv g_{YM}^2N\gg1,
\ee
where the $T^4$ scaling follows from dimensional analysis for a 4d conformal field theory, while the $\lambda$-dependence is a nontrivial result. More recently this result has been extended to include strong magnetic fields \cite{Basar:2012gh}: 
\be\label{GCSSYM2}
\Gamma_{CS}^{SYM}=\frac{\lambda^2}{384 \sqrt{3}\pi^5}\,B T^2, \ \ \ \ \ {\rm for}\ B\gg T.
\ee
In these cases, the dependence on the temperature and the magnetic field is determined by the underlying conformal invariance of the theory, but in a strongly coupled non-conformal theory there could be sizeable  deviations from this behavior. It is then well motivated to try to extend the analysis to more realistic models of holographic QCD.

In  \cite{Witten:1998zw}, Witten obtained a closer holographic analogue of QCD by wrapping D4-branes on a circle with anti-periodic boundary conditions for fermions. For small values of the effective 4d 't~Hooft coupling, $\lambda_4\ll 1$, the model behaves as 4d pure Yang-Mills theory at distances large compared to the radius $R_4$. In the opposite regime, $\lambda_4\gg 1$, the model is well-described by a weakly curved and weakly coupled 10d gravity dual up to energies that diverge with $N$  in the large-$N$ limit  \cite{Witten:1998zw}. This is the regime we will be studying. Flavors can be added to the model by introducing ``flavor branes'' in a probe approximation \cite{Kruczenski:2003uq, Sakai:2004cn}, and chiral symmetry breaking is nicely realized. In the large-$N$ limit we will be working in, flavor branes will not affect our considerations, so we will not consider them here.

In the gravity approximation, the D4-brane model with antiperiodic fermions undergoes a deconfinement phase transition at temperature $T=1/(2\pi R_4)$  \cite{Witten:1998zw}. Beyond this temperature, the dominant geometry is a black D4-brane solution with one of the spatial worldvolume dimensions compactified on a circle.

In this paper, we are interested in the Chern-Simons diffusion rate in the high temperature phase of this model. 
For the calculation we follow Son and Starinets \cite{Son:2002sd}, and we find
\be\label{CShighT}
\Gamma _{CS} = \frac{1}{2\pi}\frac{\lambda_4 ^3}{3^6\pi^2} (2\pi R_4)^2 T^6.
\ee
The $T^6$ scaling should be contrasted with the $T^4$ scaling of large-$N$, strongly coupled ${\cal N}=4$ super-Yang-Mills theory, and suggests a 6d origin. Naively, one might have expected 5d behavior instead, because the inverse temperatures we consider are small compared to $R_4$ but large compared to the M-theory circle (which scales like $1/N$ in the large-$N$ limit).  In order to try and gain some insight into this, we also consider D4-branes wrapped on a circle with susy-preserving periodicity conditions. Now we find the result \eq{CShighT} even for a range of temperatures below the Kaluza-Klein scale, which appears even more puzzling.

However, $T^6$ scaling was previously observed for the free energy of D4-branes \cite{Klebanov:1996un, Klebanov:1997kv}, which already suggested that D4-brane thermodynamics knows about a sixth dimension even at low energies compared to the scale set by the M-theory circle. The 6d behavior of D4-branes was explained from a holographic point of view in \cite{Kanitscheider:2008kd, Kanitscheider:2009as}, using a generalized conformal symmetry the D4 inherits from the conformal M5-brane theory. This has been used to derive relations between D4 and M5 hydrodynamic quantities \cite{Kanitscheider:2009as}.

We therefore look for a higher-dimensional interpretation of the Chern-Simons diffusion rate, and find that the 4d Chern-Simons diffusion rate is related to an instanton current diffusion rate in 5d and to the shear viscosity of M5-branes. This adds an entry to the D4/M5 hydrodynamic dictionary \cite{Kanitscheider:2009as}. Finally, following \cite{HoyosBadajoz:2010td} we suggest a field theory interpretation of the 6d behavior using ``deconstruction'' or ``fractionization''.

To summarize, the main results of our work are the computation of the Chern-Simons diffusion rate in the high-temperature phase of Witten's holographic 4d Yang-Mills model, an added entry in the map between D4-brane and M5-brane hydrodynamics and a suggested field theory interpretation of the 6d behavior.

The remainder of this paper is organized as follows. In section \ref{models} we review the holographic Yang-Mills models based on compactified D4 branes, with and without anti-periodic boundary conditions for fermions. In section \ref{CS} we present our calculation of the Chern-Simonds diffusion rate in  the high temperature phase.
Finally, in section \ref{6d} we discuss the result in terms of quantities in a six-dimensional conformal field theory and suggest an explanation for this behavior.

\section{Holographic models based on D4-branes on a circle}\label{models}

The models we will be interested in are based on the near horizon limit of a stack of D4 branes with one of the world-volume directions compactified on a circle, with radius $R_4$. In Witten's model \cite{Witten:1998zw}, supersymmetry is broken by imposing antiperiodic boundary conditions for the fermions. In the regime in which the gravity approximation is reliable, two classical geometries will compete in the partition function. Depending on the temperature, one or the other will dominate. At temperatures low compared to the Kaluza-Klein scale, $T<1/(2\pi R_4)$, the dominant spacetime is%
\footnote{
We use a notation similar to that of \cite{Kruczenski:2003uq, Aharony:2006da}.
}
\begin{align}\label{lowTmetric}
\notag &ds^2=\left(\frac{r}{R}\right)^{3/2}\left[-dt^2+\delta_{ij} dx^i dx^j+f(r)(dx^4) ^2\right] +\left(\frac{R}{r}\right)^{3/2}\left[\frac{dr^2}{f(r)}+r^2d\Omega_4^2\right],  \\
&F^{(4)}=\frac{2\pi N}{V_4}\,\epsilon_4,\ \ \ \ e^\phi=g_s\left(\frac{r}{R}\right)^{3/4},\ \ \ \   f(r)=1-\left(\frac{r_\Lambda}{r}\right)^3,
\end{align}
with
\be\label{R}
R^3=\pi g_s N \ell_s^3.
\ee
The coordinates $r$ and $x^4$ span a cigar geometry, which is non-singular if $x^4$ is periodically identified with period
\be\label{rLambda}
2\pi R_4=\frac{4\pi}{3}\left(\frac{R^3}{r_\Lambda}\right)^{1/2}.
\ee
In \eq{lowTmetric}, $V_4=8\pi^2/3$ is the volume of the unit 4-sphere and $\epsilon_4$ the corresponding volume element. The 4d 't~Hooft coupling at the compactification scale is
\be\label{lambda4}
\lambda_4=g_4^2N=\frac{2\pi g_s\ell_sN}{R_4}.
\ee
The typical masses of glueballs in this model are of order $1/R_4$, whereas the QCD string tension is of order $\lambda_4/R_4^2$. The geometry \eq{lowTmetric} is weakly curved in the regime $\lambda_4\gg 1$, and the dilaton is small as long as we consider energies small compared to an energy scale that scales like $N^{4/3}$ in the 't~Hooft large-$N$ limit (which we will do in this paper; at higher energies, an M-theoretic description in terms of M5-branes would have to be used). Note that in the gravity regime the Kaluza-Klein scale is comparable to the mass gap, so that the dynamics is not really four-dimensional. On the other hand, in the regime  $\lambda_4\ll 1$, the field theory is weakly coupled  and the low-energy modes are described by 4d pure Yang-Mills theory. 

In the high-temperature regime, $T>1/(2\pi R_4)$, the dominant spacetime is the near-horizon geometry of near-extremal D4-branes:
\begin{align}\label{highTmetric}
\notag &ds^2=\left(\frac{r}{R}\right)^{3/2}\left[- f(r)dt^2+\delta_{ij} dx^i dx^j+(dx^4) ^2\right] +\left(\frac{R}{r}\right)^{3/2}\left[\frac{dr^2}{f(r)}+r^2d\Omega_4^2\right],  \\
&F^{(4)}=\frac{2\pi N}{V_4}\,\epsilon_4,\ \ \ \ e^\phi=g_s\left(\frac{r}{R}\right)^{3/4},\ \ \ \ f(r)=1-\left(\frac{r_T}{r}\right)^3.
\end{align}
Now the Kaluza-Klein radius $R_4$ is arbitrary, but the temperature is fixed by regularity of the Wick-rotated geometry ($t=-i\tau$):
\be\label{rT}
\frac{1}{T}=\frac{4\pi}{3}\left(\frac{R^3}{r_T}\right)^{1/2}.
\ee
Outside the horizon, in terms of the 5d 't~Hooft coupling,
\be
\lambda_5 \equiv (2\pi)^2g_s\ell_s N,
\ee
the spacetime \eq{highTmetric} is weakly curved  if $T\gg 1/\lambda_5$ and $\lambda_5 \gg \ell_s$. The length of the circle along $x^4$ is large compared to the string length if $T\gg 1/(R_4^{2/3}\lambda_5^{1/3})$. Finally, the string coupling is small if $T\ll N^{2/3}/\lambda_5$.

One assumes/hopes that the phase transition between the geometries \eq{lowTmetric} and \eq{highTmetric} (at $\lambda_4\gg 1$) is related to the deconfinement phase transition of large-$N$ Yang-Mills theory (at $\lambda_4\ll 1$). However, as discussed in \cite{Aharony:2006da, Mandal:2011ws}, the connection cannot be a direct continuation because the center symmetry in the spatial direction is broken in the deconfined phase but not in the phase described by the black D4 brane.

Later on, we will also be interested in the supersymmetric model with periodic boundary conditions along $x^4$ on the fermions. In this case, the spacetime \eq{lowTmetric} is not allowed, and the spacetime  \eq{highTmetric} always dominates when it is a reliable supergravity solution. It follows from the discussion above that if $\lambda_5\gg R_4$, the regime of validity of the supergravity approximation includes a range of temperatures below the Kaluza-Klein scale, 
\be\label{Trange}
1/(R_4^{2/3}\lambda_5^{1/3})\ll T < 1/R_4, 
\ee
where one might expect 4d physics.

\section{Computing the Chern-Simons diffusion rate}\label{CS}

Upon dimensional reduction, the worldvolume action of a flat D4-brane wrapped around a circle with radius $R_4$ contains the terms
\be
S_{D4} \supset -\left(\frac{R_4}{2\pi g_s\ell_s}\right)\int d^4x\,\frac{1}{4} F_{ab} F^{ab}\,+\frac{1}{64\pi^2}\int d^4x\left(\frac{4\pi R_4 C^{(1)}_4}{g_s\ell_s}\right)\epsilon^{abcd}F_{ab}F_{cd},
\ee 
where from an AdS/CFT point of view the first term justifies the identification \eq{lambda4}, while the second term implies that the asymptotic behavior of the Ramond-Ramond one-form component along the circle corresponds to the axion (whose constant mode plays the role of a theta angle) \cite{Witten:1998uka},
\be\label{thetaC}
\theta=\frac{4\pi R_4 C^{(1)}_4}{g_s\ell_s}.
\ee

In terms of its field strength $F^{(2)}_{\mu\nu}=\partial_\mu C^{(1)}_\nu - \partial_\nu C^{(1)}_\mu$ and the spacetime metric $G_{\mu\nu}$, the kinetic term of the Ramond-Ramond one-form is
\be\label{kinetic}
-\frac{1}{256 \pi^7 g_s^2 \ell_s^8}\int d^{10}x\sqrt{-G}\frac12 G^{\mu\nu} G^{\rho\sigma} F^{(2)}_{\mu\rho} F^{(2)}_{\nu\sigma}.
\ee

In order to calculate the Chern-Simons diffusion rate in the high-temperature phase, we apply the method developed in \cite{Son:2002sd} to the geometry \eq{highTmetric}, to which we apply a coordinate transformation $u=r_{T}/r$:
\begin{equation}
ds^{2} = \left(\frac{r_{T}}{R u}\right)^{3/2}\left[-(1-u^{3})dt^{2}+\delta_{ij} dx^i dx^j+(dx^4)^{2}\right]
 +R^{3/2}r_{T}^{1/2}\left[\frac{du^{2}}{u^{5/2}(1-u^{3})} +\frac{1}{u^{1/2}}d\Omega_4^2\right],
\end{equation}
so that the boundary and horizon values of $u$ are $u_B=0$ and $u_H=1$, respectively. In this geometry, the kinetic term \eq{kinetic} for the component $C^{(1)}_4$ reduced to 5d takes the form
\bea\label{action}
&&-\frac{R_4 r_T^3}{48\pi^4g_s^2\ell_s^8}\int d^4x\, du \left[\frac{1-u^3}{u^2}\,\partial_u C_4^{(1)}\partial_u C_4^{(1)} -\frac{R^3}{r_T u^3(1-u^3)}\,\partial_t C_4^{(1)}\partial_t C_4^{(1)}\right. \nonumber\\
&&\left.\ \ \ \ \ \ \ \ \ \ \ \ \ \ \ \ \ \ \ \ \ \ \ \ \ \ \ \ + \frac{R^3}{r_T u^3}\,\partial_i C_4^{(1)}\partial_i C_4^{(1)} \right],
\eea
leading to the equation of motion
\be\label{eom}
\frac{r_T(1-u^3)}{ u^2}\partial_u^2 C_4^{(1)} -\frac{r_T(2+u^3)}{ u^3}\partial_u C_4^{(1)} -\frac{R^3}{ u^3(1-u^3)}\partial_t^2 C_4^{(1)} +\frac{R^3}{ u^3}\partial_i^2 C_4^{(1)}=0.
\ee
Following  \cite{Son:2002sd}, and taking into account the relative normalization \eq{thetaC} between $C^{(1)}_4$ and $\theta$, we Fourier transform
\begin{equation}\label{Fourier}
C_4^{(1)} =\frac{g_s\ell_s}{4\pi R_4}\int \frac{d^4 k}{(2\pi)^4} e ^{i k\cdot x}f_k (u) \theta_0(k),
\end{equation}
where the modes $f_k$ satisfy incoming-wave boundary conditions at the horizon, and approach 1 at the boundary, $f_k(u_B)=1$. The coefficients $\theta_0(k)$ are then determined by the boundary profile of $C^{(1)}_4$:
\begin{equation}
C_4^{(1)}(u_B,x) =\frac{g_s\ell_s}{4\pi R_4}\int \frac{d^4 k}{(2\pi) ^4} e ^{i k\cdot x}\theta_0(k).
\end{equation}
On-shell, the action then reduces to the surface terms
\begin{equation}
S_{\rm on-shell}=\int\frac{d^{4}k}{(2\pi)^{4}}\theta_0(-k)\left.\mathcal{F}(k,u)\theta_0(k)\right|^{u=u_{B}}_{u=u_{H}},
\end{equation}
where
\begin{equation}\label{calF}
\mathcal{F}(k,u)=\frac{r_T^3}{2^8\,3\pi^6\ell _s^6R_4}\frac{1-u^3}{u^2}f_{-k}(u)\partial_{u}f_{k}(u).
\end{equation}
According to Son and Starinets \cite{Son:2002sd}, the retarded Green's function for $(1/(64\pi^2))\epsilon^{\mu\nu\rho\sigma}\, F^a_{\mu\nu} F^a_{\rho\sigma}$, the operator sourced by the rescaled Ramond-Ramond potential \eq{thetaC}, is given by
\begin{equation}\label{GR}
G^R (k)=-2 \mathcal{F}(k,u)|_{u=u_B=0}.
\end{equation}
From this retarded propagator, the Chern-Simons diffusion rate can be computed using the Kubo formula
\begin{equation}\label{Kubo}
\Gamma _{CS} =  -\lim_{\omega\rightarrow0}\frac{2T}{\omega}\mathrm{Im}\, G^{R}(\omega,\vec 0).
\end{equation}
To complete the computation of the Chern-Simons diffusion rate, we therefore need to determine the modes $f_k(u)$, at least for vanishing spatial momenta $\vec k$ and small frequency $\omega$. Up to an overall normalization factor that goes to 1 in the limit of small frequencies and momenta, the solution to the mode equation following from \eq{eom} with the boundary conditions mentioned after \eq{Fourier} is
\be
f_k(u)=(1-u)^{-\frac{i\omega}{4\pi T}}\left[1-\frac{i\omega}{4\pi T}\ln(1+u+u^2)\right] + O(\omega^2,\vec k^2)=1-\frac{i\omega}{4\pi T}\ln(1-u^3) + O(\omega^2,\vec k^2).
\ee
Using \eq{calF} and \eq{GR}, this leads to
\be
G^R (k)=-\frac{i\omega r_T^3}{2^9\pi^7R_4\ell_s^6T}+ O(\omega^2,\vec k^2),
\ee 
after which \eq{Kubo}, \eq{R}, \eq{lambda4} and \eq{rT} lead us to the following result for the Chern-Simons diffusion rate in the high-temperature phase:
\begin{equation}\label{GCSresult}
\Gamma _{CS} = \frac{2 \lambda_4 ^3 R_4^2 T^6}{3^6 \pi}.
\end{equation}
In the supersymmetric model with periodic fermions (in which the ``low-temperature'' bulk geometry does not exist), this result is also valid in the temperature range \eq{Trange}.

\section{Relation to six dimensions}\label{6d}

The $T^6$ scaling of \eq{GCSresult} contrasts with the $T^4$ scaling of \eq{GCSSYM}, which was guaranteed by 4d conformal invariance. One may wonder whether it reflects an underlying 6d conformal field theory, namely that of the M-theory M5-branes the D4-branes descend from. Naively, it sounds strange that 6d behavior would be visible at temperatures small compared to the inverse size of the M-theory circle (whose radius scales like $1/N$ in the 't~Hooft limit, so it will always be small for our considerations). In the supersymmetric version of the model, the $T^6$ scaling of the Chern-Simons diffusion rate persists even  in the temperature range \eq{Trange}, where the system would seem to be clearly four-dimensional, which appears even more puzzling. In this section, we attempt to gain some intuition for this behavior.

\subsection{Free energy}

A first hint comes form studying the entropy density of D4-branes wrapped on a circle. Again, at high temperatures one might have naively expected the simplest possible behavior in a five-dimensional theory, namely
\begin{equation}
s\sim N^2 T^4,
\end{equation}
while in the supersymmetric model one might have expected that at low temperatures the entropy is similar to a four-dimensional theory, with corrections coming from the massive Kaluza-Klein modes on the fifth direction
\begin{equation}
s\sim N^2T^3\left(1+O(T e^{-1/(R_4T)})\right).
\end{equation}
However, this is not the case: at any temperature the 5d entropy density is \cite{Klebanov:1996un, Klebanov:1997kv}
\begin{equation}
s_{D4}=\frac{2^6\pi^2}{3^6}g_5^2 N^3 T^5; \ \ \ \ \ \ \ \ \ g_5^2\equiv 4\pi^2g_s\ell_s.
\end{equation}
This coincides with the entropy of the black M5 brane compactified on a circle of radius $R_{11}$, as already pointed out in \cite{Klebanov:1997kv}:
\begin{equation}
s_{D4}=\frac{2^6\pi^2}{3^6}g_{5}^2 N^3 T^5=(2\pi R_{11})\frac{2^7\pi^3}{3^6}N^3 T^5=(2\pi R_{11})s_{M5},
\end{equation}
where the value of $s_{M5}$ was computed in \cite{Klebanov:1996un}.

The coincidence between the entropies suggests that the finite temperature theory knows about its six-dimensional origin, no matter how small the temperature is compared to the inverse of the compactification radius. This point of view is reinforced by the analysis of \cite{Kanitscheider:2008kd}, mapping the energy-momentum tensor of the two theories, which was subsequently used to derive the hydrodynamics of the D4 theory from the M5 theory \cite{Kanitscheider:2009as}. We will now show that our result on the Chern-Simons diffusion rate indeed descends from 6d hydrodynamics and provides an additional entry in the D4/M5 hydrodynamics correspondence explored in \cite{Kanitscheider:2009as}.

\subsection{Conductivity of the instanton current and M5 shear viscosity}

In \cite{Kanitscheider:2008kd} Kanitscheider, Skenderis and Taylor derived the holographic D4 energy-momentum tensor from the M5 theory using holographic renormalization. They showed that the components of the energy-momentum tensor and the ``gluon condensate'' (the operator dual to the dilaton) in the D4 and M5 theory are simply related to components of the M5 energy-momentum tensor
\begin{align}
\notag &R_{11}\vev{T_{MN}}^{D4}\sim \vev{T_{MN}}^{M5}, \ \ M,N=0,1,2,3,4.\\
&R_{11}\vev{\cal O_\phi} \sim \vev{T_{55}}^{M5}.
\end{align}
In \cite{Kanitscheider:2009as} this was used to derive relations between thermodynamic quantities, like the relation between the entropies we have just discussed, and relations between transport coefficients.\footnote{The analysis there was not limited to the D4 but we will discuss only this case here.} For instance, from the reduction one can show that the speed of sound of the D4 theory is the same as in the M5, $c_s^2=1/5$, and that the bulk over shear viscosity ratio obeys the bound
\begin{equation}
\frac{\zeta_{D4}}{\eta_{D4}}\geq 2\left(\frac{1}{4}-c_s^2 \right).
\end{equation}

Let us therefore interpret the Chern-Simons diffusion rate of D4-brane models on a circle in terms of M-theory. Upon lifting to M-theory, the component $C^{(1)}_4$ is related to the metric component $G^{(11)}_{45}$, with the index 5 labeling the M-theory circle. The D4-brane instanton current
\begin{equation}
J={}^{*_5} \tr(F\wedge F)
\end{equation}
lifts to a current of momentum along the M-theory circle; its component $J_4$ corresponds to the stress-tensor component $T^{M5}_{45}$. When the D4-branes are compactified along the $x^4$ direction, the $J_4$ component becomes the topological charge density of the 4d theory, $\tr(F\wedge F)$.  The Chern-Simons diffusion rate should therefore be related to the shear viscosity of M5-branes. This shear viscosity can be computed using the Kubo formula
\begin{equation}\label{KuboM5}
\eta=-\lim_{\omega\to 0}\frac{1}{\omega} G_{45,45}(\omega,\vec{q}=0),
\end{equation}
where $G_{45;45}(\omega,\vec{q})$ is the Fourier transform of the retarded correlator of the $T_{45}^{M5}$ components of the energy-momentum tensor on the M5. Using the AdS/CFT correspondence, its value was computed in \cite{Herzog:2002fn},
\begin{equation}\label{etaresult}
\eta=\frac{s_{M5}}{4\pi}=\frac{2^5\pi^2}{3^6}N^3 T^5.
\end{equation}
We now show how to reproduce our Chern-Simons diffusion rate \eq{GCSresult} starting from the shear viscosity \eq{etaresult} of the M5 theory. First, note that the Kubo formula \eq{KuboM5} does not contain the factor $2T$ present in \eq{Kubo}; so let us multiply \eq{etaresult} by $2T$. Next, multiply by $(g_s\ell_s)^2/(4\pi R_4)^2$ to account for the relative normalization \eq{thetaC} between $\theta$ and $C^{(1)}_4$. Then convert the 6d density in a 4d density by multiplying by the lengths of the two compactification circles, $2\pi g_s\ell_s$ and $2\pi R_4$. Finally, use \eq{lambda4}. The result of these manipulations is exactly \eq{GCSresult}.

\subsection{Relations to the spectrum of M5-branes on a torus}

The fact that the M5 and D4 brane have the same thermodynamics is relatively straightforward from the perspective of the gravity dual: the D4 geometry can be obtained applying dimensional reduction on the M5 geometry and therefore their action is the same. As we have discussed before, from the field theory perspective the relation is more mysterious, especially at temperatures small compared to the inverse radii of compactification $1/R_4$ and $1/R_{11}\equiv1/(g_s\ell_s)$. What makes the D4 brane look six-dimensional even at such temperatures?

We will argue below that the Kaluza-Klein spectrum of a collection of $N$ M5-branes is not set by the compactification radii $R_4$ and $R_{11}$, but by effective compactification radii $NR_4$ and $NR_{11}$. Because we are working in a regime in which $R_{11}\ll R_4$, this would imply that 6d behavior would be expected for $T \gg 1/(NR_{11})$. This would match perfectly with the fact that \eq{highTmetric} is weakly curved outside the horizon precisely for such temperatures. For lower temperatures, the gravity approximation is not valid and there is no reason to expect 6d behavior. So what remains to be shown is why the effective Kaluza-Klein radii are $N$ times larger than the naive ones. We will give two (related) arguments.

The first argument relies on the concept of ``fractionization'' used in the context of counting black hole microstates; see, for instance, \cite{Mathur:2005ai} for a review. It is well-known that $N$ fundamental strings wrapping a circle of radius $R$ can reconnect so as to become one fundamental string wrapping the circle $N$ times. The total length of the string is then $2\pi NR$, and the smallest non-zero momentem the string can carry is $1/(NR)$ rather than $1/R$. Dualizing the $N$ fundamental strings to $N$ M5-branes, one finds that the latter can also reconnect and carry momenta $N$ times smaller than one would naively have thought. A similar situation can be found in more ordinary conformal field theories at large-$N$, and as explained in \cite{Unsal:2010qh,Poppitz:2010bt} this leads to the independence of some quantities of the volume of compactification (See \cite{Mandal:2011ws} for a related discussion in the present context). Perhaps similar arguments can be generalized to the M5 theory in the regime described by supergravity.

The second argument, which we will describe in the rest of this section, uses ``deconstruction''. First apply T-duality along the $x^4$ direction in the D4 geometry. At the level of supergravity one simply apply Buscher's rules, which in this particular case take the simple form
\begin{equation}
g_{44}'=1/g_{44}, \ \ e^{2\phi'}=e^{2\phi}/g_{44}, \ \ C_{0123}'=C_{01234}.
\end{equation}
The resulting geometry is dual to a smeared configuration of $N$ D3 branes. Note that {\em at the level of supergravity} one cannot get a localized D3 brane configuration, since the isometry along the $x^4$ direction is preserved by the T-duality transformation. The length of the $x^4$ direction is $\ell_s^2/(2\pi R_4)$ in the D3 geometry. 

The same geometry as the smeared $N$ D3 branes can be obtained by considering a uniform (evenly spaced)  distribution of D3 branes along the $x^4$ direction. The separation between neighbouring D3 branes is $\sim \ell_s^2/(N R_4)$, and when $N$ is very large it cannot be resolved at the level of supergravity. Therefore, the leading large-$N$ thermodynamics of the D4 branes will be the same as for this uniform distribution. From the point of view of the $\cN=4$ $SU(N)$ super Yang-Mills theory that describes the D3 branes at low energies, their relative separation translates into expectation values of the scalar fields belonging to the vector supermultiplet. Therefore, the uniform distribution corresponds to a particular vacuum in the $\cN=4$ Coulomb branch. 

Note that if we take as a starting point a uniform distribution of D3 branes along the $x^4$ direction and apply T-duality, the resulting configuration is a D4 brane with a non-trivial Wilson line along the $x^4$ direction, which will also produce some Higgsing in the D4 gauge theory. 

An explanation for the six-dimensional behavior of the theory for the smeared D3 geometry is based on deconstruction: the low-energy dynamics of a distribution of D3 branes on a circle reproduces the M5 theory, as was shown in \cite{ArkaniHamed:2001ie} for the Higgs branch of an orbifold of $\cN=4$ super Yang-Mills, and at low energies for a Coulomb branch configuration with a distribution of eigenvalues on a line in \cite{HoyosBadajoz:2010td}, on which we base the argument below.

The low-energy spectrum consists of $N$ $\cN=4$ $U(1)$ massless vector multiplets, one for each D3 brane, and supermultiplets of W bosons whose mass is generated by the separation of the D3 branes that produces the breaking $SU(N)\to U(1)^{N-1}$. The mass of the W bosons is proportional to the separation between D3 branes times the string tension. Since we assume that D3 branes are evenly separated, the mass spectrum of W bosons will take the form
\begin{equation}
m_W\sim \frac{n}{N R_4} \ \ \ n=1,2,\cdots
\end{equation}
For $n=1$, $m_W$ corresponds to the modes between two neighbouring D3 branes. For D3 branes set further apart the mass of the associated W bosons increases with the distance. Note that the resulting spectrum can also be interpreted as a tower of Kaluza-Klein modes on a circle of radius
\begin{equation}
\tilde{R}_4\sim NR_4.
\end{equation}
From the $SL(2,\mathbb{Z})$ invariance of the $\cN=4$ theory, one can deduce that there should be similar tower of states for magnetically charged objects, forming in total a Kaluza-Klein tower of modes on a tours. The mass of the purely magnetically charged states, which correspond to stretched D1-branes, is
\begin{equation}
m_{\widetilde W}\sim \frac{n}{g^2_{YM} N R_4}\sim \frac{n}{\lambda_4 R_4} , \ \ \ n=1,2,\cdots
\end{equation}
so the radius of the second direction in the torus will be
\begin{equation}
\tilde{R}_5\sim \lambda_4R_4\sim NR_{11}.
\end{equation}
This is what we set out to show. This system has 16 supercharges, because it is a vacuum state of $\cN=4$ SYM theory in four dimensions. The $(2,0)$ theory that describes the low-energy physics of the M5 brane has a similar spectrum to the 1/2 BPS states described by the $W$ bosons and their dyonic versions.

A deconstruction argument can also be formulated directly in the D4 field theory. In this case, the tower of massive states will be generated by the breaking of the gauge group $SU(N)\to U(1)^{N-1}$ by a Wilson line along the $x^4$ direction. There is no $SL(2,\mathbb{Z})$ symmetry, so  only one additional direction will be generated, resulting again in six-dimensional behavior.


\section*{Acknowledgments}

We would like to thank T.~Morita for clarifying comments on the phases of the black D4 brane, and U.~G\"ursoy and A.~O'Bannon for pointing out an incorrect statement in the published version of this paper.
This research was supported in part by the Belgian Federal Science Policy Office through the Interuniversity Attraction
Pole IAP VI/11, by FWO-Vlaanderen through project G011410N, by the Israel Science Foundation (grant number 1468/06) and by the National Science Foundation under Grant No.\ PHY11-25915.
BC is grateful to the Newton Institute, the KITP and the University of Amsterdam for hospitality and partial support while this work was in progress. 
PS is a Postdoctoral Researcher of FWO-Vlaanderen, and would like to thank University of Witwatersrand and Oxford University for hospitality and partial support during the course of this work. 


\end{document}